\documentclass[runningheads]{llncs}

\usepackage{graphicx}
\usepackage{tabularx}
\usepackage{longtable}
\usepackage{makecell}
\usepackage{multirow}
\usepackage{multicol}
\usepackage{arydshln}
\usepackage{booktabs}
\usepackage{xcolor}
\usepackage{lscape}
\usepackage{dblfloatfix}  
\usepackage{dblfloatfix}   

\usepackage{multicol}
\usepackage{enumitem}

\newcommand\blfootnote[1]{%
    \begingroup
    \renewcommand\thefootnote{}\footnote{#1}%
    \addtocounter{footnote}{-1}%
    \endgroup
}
\begin{document}

\title{Social Companion Robots to Reduce Isolation: A Perception Change Due to COVID-19}%

%
%



\author{Moojan Ghafurian\inst{1} \and
Colin Ellard\inst{2} \and
Kerstin Dautenhahn\inst{1,3}}
\authorrunning{Ghafurian et al.}
%
\institute{Department of Electrical and Computer Engineering \and
Department of Psychology
 \and
Department of Systems Design Engineering\\
University of Waterloo, Waterloo, Ontario, Canada\\
\email{\{moojan, cellard, kerstin.dautenhahn\}@uwaterloo.ca}}

\maketitle              
\begin{abstract}
Social isolation is one of the negative consequences of a pandemic like COVID-19.  Social isolation and loneliness are not only experienced by older adults, but also by younger people who live alone and cannot communicate with others or get involved in social situations as they used to. In such situations, social companion robots might have the potential to reduce social isolation and increase well-being. However, society's perception of social robots has not always been positive. In this paper, we conducted two online experiments with 102 and 132 participants during the self isolation periods of COVID-19 (May-June 2020 and January 2021), to study how COVID-19 has affected people's perception of the benefits of a social robot. Our results showed that a change caused by COVID-19, as well as having an older relative who lived alone or at a care center during the pandemic significantly and positively affected people’s perception of social robots, as companions, and that the feeling of loneliness can drive the purchase of a social robot. The second study replicated the results of the first study. We also discuss the effects of Big 5 personality traits on the likelihood to purchase a social robot, as well as on participants' general attitude towards COVID-19 and adapting to the pandemic.\blfootnote{Note - Cite as: Ghafurian, M., Ellard, C., \& Dautenhahn, K. (2021, August). Social companion robots to reduce isolation: A perception change due to COVID-19. In IFIP Conference on Human-Computer Interaction (pp. 43-63). Springer, Cham.}
\end{abstract}

\section{Introduction}


Social isolation is a common issue among older adults, especially those with disabilities, and can affect health and quality of life. Lack of social support can lead to loneliness, which is a significant public health issue~\cite{gerst2015loneliness} and is associated with multiple negative health outcomes, such as depression~\cite{alpass2003loneliness,bennett2017robot}, anxiety~\cite{santini2020social}, and an increase in cardiovascular risk~\cite{smith2002psychosocial}. While social isolation is more common among older adults, pandemics such as COVID-19 can lead to an increase in social isolation not only in older adults~\cite{armitage2020covid}, but among adults of all ages~\cite{usher2020life,douglas2020mitigating}. Social isolation is an important consequence of COVID-19, along with many other negative consequences~\cite{douglas2020mitigating} such as an increase in family violence~\cite{usher2020family}. As concluded by Van Bavel et al.~\cite{van2020using}, action is required to mitigate these devastating effects.

Since the outbreak of COVID-19, many researchers have emphasized different benefits of using robots and artificially intelligent systems on mitigating the associated risks. These applications included using robots for medical education~\cite{goh2020vision}, managing the spread of COVID-19 in public areas by reducing human contact, disinfecting areas,  providing security~\cite{zeng2020high,yang2020combating}, and helping children to stay connected with each other \cite{Scassellatieabc9014}. We believe that another important application area is to use social robots as companions during a pandemic such as COVID-19. Social robots have the potential to act as companions and reduce social isolation, and have already been used successfully  in providing care and support in many domains, such as by improving older adults' mood~\cite{wada2005psychological,wada2002robot}, reducing depression~\cite{abdollahi2017pilot}, and even reducing the need to use medication \cite{shibata2012therapeutic}.

However, perception of social robots has not always been positive, and previously, prior to COVID-19, it has been reported that there is an increasing trend in people's negative attitudes towards robots~\cite{gnambs2019robots}. It might also be difficult to perceive their benefits accurately, especially by those who do not have any experience of, or direct knowledge of anyone who has experienced loneliness and social isolation. One of the common concerns raised by many researchers and participants alike is that social robots may replace human companions. While this is a valid concern, social robots can be highly beneficial in situations where such human interaction is in fact prohibited, for example during a pandemic such as COVID-19, or for older adults who might be socially isolated, e.g., due to medical conditions. In these situations the presence of social robots could be positive and effective in reducing their loneliness and improving society's mental wellbeing. 

While companion robots can be beneficial to reduce loneliness, they can only succeed if society has a positive attitude towards them. Specifically, people need to perceive the benefits of social robots and be willing to adopt them. It has been previously shown that lower social support can increase older adults' acceptance of social robots that are less intuitive to use~\cite{baisch2017acceptance}. But, a lower level of psychological well-being, such as emotional loneliness, life satisfaction, and depressive mood can reduce acceptance, depending on the robot and how intuitive it is to use~\cite{baisch2017acceptance}. We argue that the changes caused by COVID-19 has resulted in a higher level of attention to the consequences of social isolation, which could highlight the benefits of having a companion robot (for self or loved ones), and could change society's perception of them.

Therefore, we investigated how experience of social isolation due to COVID-19 has affected the perception of benefits of companion robots, and whether loneliness can affect the tendency to purchase a social robot (as an indicator of participants' intention to use it). We further asked what tasks people would prefer for a companion robot to carry out, and what aspects of a companion robot they perceived to be important.

\subsection*{Research Questions and Hypotheses:} 

This paper investigated seven research questions:

\begin{itemize}
    \item [\bf RQ1:] Has a change due to COVID-19 affected people's perception of social companion robots?
    \item [\bf RQ2:] Has having an older relative who lives alone or at a care center during the pandemic affected people's perception of social companion robots?
    \item [\bf RQ3:] Can loneliness increase the likelihood of purchasing a companion robot?
    \item [\bf RQ4:] Do people have strong preferences for tasks of companion robots?
    \item [\bf RQ5:] Are social elements (e.g., showing emotions) perceived to be important for a companion robot, and how does the importance of such components compare with the technical accuracy of the robots?
    \item [\bf RQ6:] How do different personality types affect attitudes towards COVID-19, as well as attitudes towards social robots during the pandemic?  
    \item [\bf RQ7:] Are the findings related to the perception of social robots due to COVID-19 and the effect of loneliness on purchase of a social robot robust and independent of the time that the study was conducted?  
\end{itemize}

\noindent Our hypotheses were as follows.

\begin{itemize}
    \item [\bf H1:] COVID-19 has caused people to reflect more about the consequences of social isolation, and has positively affected attitudes towards social companion robots, because the pandemic pointed out situations where social robots can fill gaps in the provision of social interaction, as opposed to robots being perceived negatively as replacements of human contact.
    \item [\bf H2:] For the same reason as in H1, having an older relative who either lives alone or at a care center has positively affected the attitude towards social robots. 
    \item [\bf H3:] A change in perception of social robots can positively influence people's tendency to purchase one.
    \item [\bf H4:] Loneliness can be an important factor in adopting a social robot and can increase the tendency to purchase a social companion robot.
    \item [\bf H5:] Social elements such as robots' ability to show emotions, adapting its behaviour based on its users, etc., will be perceived to be as important as technical accuracy by people, since many studies have shown their advantages in experimental settings.  
    \item [\bf H6:] The observed effects can be replicated at a later time, as we believe that these effects (e.g., change of perception of social robots due to COVID-19) are long-term, as a result of COVID-19, and independent of the time of the study.
\end{itemize}

\section{Background}

In previous studies, social robots were successfully used in many contexts, such as for increasing older adults' social engagement~\cite{vsabanovic2013paro,perugia2017modelling} and providing companionship~\cite{odetti2007preliminary,mannion2019introducing}. They also helped with activities of daily living, such as helping older adults with dementia with the process of eating food~\cite{derek2012socially}, as well as supporting nurses in a hospital~\cite{van2019social}. Furthermore, social robots have been able to help children with autism in many domains such as therapy and education~\cite{dautenhahn2003roles,robins2009isolation}, for example by increasing their social engagement~\cite{robins2005robotic}, or even involving children in 
peer-learning scenarios, e.g., to improve children's writing skills~\cite{chandra2019children}. 

However, despite their success in multiple domains and a variety of studies, many of the existing social robots have not been used much beyond the context of these studies. A part of this can be due to the existing challenges, such as having robots that can act fully autonomously, or concerns related to a robot's monetary cost~\cite{moyle2014connecting,picking2017exploring} (as it is challenging to build affordable consumer robots~\cite{share2018preparing}). 

To be successfully adopted by their intended users, multiple factors are critical, which can be related to the robot (e.g., its appearance and design, its capabilities, etc.) and its users (e.g., users' acceptance, trust, attitude towards robots, and likelihood of adopting a robot). User acceptance is in fact essential for the success of social robots~\cite{share2018preparing}.

According to a recent study conducted in Europe~\cite{gnambs2019robots}, people have become more cautious in using robots, and their attitude towards autonomous robots has become more negative during the period 2012 to 2017. This study also showed that people are more comfortable with robots at workplaces, as opposed to the robots used in healthcare and those that are designed to help older adults. According to this study, participants' age, gender, and education level, and employment status could all affect one's attitude towards robots: men were found to have a more positive attitude towards robots and education was positively correlated with a positive attitude~\cite{gnambs2019robots}. The growing anxiety in using robots was associated with the concern that the robots may replace humans and lead to losing one's job~\cite{nanevasystematic,broadbent2012attitudes,manyika2017jobs}. A negative attitude towards social robots can not only affect their acceptance~\cite{robb2020robots} and adoption (as people with a highly negative attitude tend to avoid human–robot communication~\cite{nomura2006experimental}), but also can affect people's interactions with the robot, and as shown by Nomura et al., users' self-expression towards the robots~\cite{nomura2006experimental}.

In general, society's perception and acceptance of social robots are affected by multiple factors. For example, de Graaf et al. recommended creating a clear purpose for robots, increasing robots' social capabilities, and considering the use context (e.g., living situation, time and location of use) to be important for developing social robots that are accepted by  society~\cite{de2016long}. In the context of healthcare, many factors such as a robot's design, personality, adaptability, humanness, size, and gender can affect how people react to it and affect users' acceptance~\cite{broadbent2009acceptance}. Furthermore, it has been shown that the attitude towards social robots can significantly affect their acceptance~\cite{nomura2009influences}. 

There are multiple other factors that have been shown to affect acceptance of social robots, including but not limited to, feelings of social presence~\cite{heerink2008influence}, perceived enjoyment~\cite{heerink2008influence}, familiarity with robots~\cite{bishop2019social}, robots being functionally relevant~\cite{de2017they}, emotional displays (e.g., positive versus negative)~\cite{bishop2019social}, robots' ease of use~\cite{de2017they}, level of interaction between potential users and robots~\cite{paetzel2020persistence}, the category label assigned to the robot (i.e., the way the robot is introduced, e.g., a "home appliance")~\cite{kim2020s}, and even age~\cite{robb2020robots}. 

Furthermore, previous studies showed different results regarding the impact of loneliness on the  acceptance of robots: loneliness has been shown to be correlated with a more positive attitude towards robots (by increasing the perceived social presence)~\cite{lee2006physically} and a more negative attitude (by decreasing anthropomorphic tendencies)~\cite{li2020does} towards social robots. However, in many cases, social robots have been promising in reducing loneliness and social isolation, especially in older adults~\cite{poscia2018interventions}.
\section{Experiment}

To address our research questions, two online studies were conducted in May 2020 and January 2021 in Canada, both during the self isolation period of COVID-19. While both were conducted during the times that people have been self isolating, the time difference enables us to ensure that the findings are robust and independent of participants' level of experience with COVID-19.

\subsection{Methodology}

A questionnaire was designed and administered online on Amazon Mechanical Turk. The questionnaire measured different aspects of people's experience of COVID-19 and perception of robots, and was implemented in 5 sections as below:

\noindent \textbf{Section 1 - Loneliness:} Feeling of loneliness as measured through the 8-item UCLA Loneliness Scale Questionnaire (ULS-8)~\cite{hays1987short}

\noindent \textbf{Section 2 - Big 5 Personality:} Big 5 personality, i.e., personality measured through five different traits of Extroversion, Openness to experience, Conscientiousness, Extraversion, Agreeableness, and Emotionality (or Neuroticism)  was measured using the TIPI Questionnaire~\cite{gosling2003very} 

\noindent \textbf{Section 3 - Extroversion:} The specific extroversion dimension of Big 5 was assessed via 24 items related to Extroversion from the 120-item Big 5 questionnaire (IPIP-120 Personality Test)~\cite{johnson2014measuring}, to ensure consistency and accuracy in the results of the Big 5 personality test. Note that the results of this questionnaire were used only for ensuring the consistency between TIPI and IPIP-120 results, as we preferred to avoid using all items from IPIP-120 (to avoid participant fatigue).

\noindent \textbf{Section 4 - Perception of Social Robots:} A questionnaire was designed, asking participants:
\begin{itemize}
    \item [\bf(a)] whether they were likely to purchase a social robot if it is ``affordable" for them (rated on a continuous scale from ``not at all" to ``very likely").
    \item [\bf (b)] if their perception of social robots as companions has changed due to COVID-19 (rated on a continuous scale from ``Not Changed at All" to ``Completely Changed"), with a follow up question asking them to explain why it has/has not changed (answer provided as a text entry). 
     \item [\bf (c)] what task/tasks they preferred for a companion robot during COVID-19. The choices were: music/video, chitchat and joking, dancing, exercises, reminders (health related, e.g., medicine and appointments, or daily life/socially related), cooking, games, relaxation or meditation or breathing exercises, storytelling or reading together, and Other (we asked them to specify).
     \item [\bf (d)] what appearance they preferred a social robot to have. The choices were: Human-like, Animal-like, and Other (we asked them to specify).
     \item [\bf (e)] what were the elements in a social robot that were important to them. The choices were: Not making mistakes, Ability to show emotions, Not requiring much maintenance, Having a specific behaviour that they might like, Recognizing them, and Other (please specify). 
\end{itemize}

All questions that were not rated on a continuous scale (i.e., preference about the robot's appearance, important elements for a social robot, and preferred tasks) were provided as multiple-choice questions, where participants could select as many responses as they wished, with the response of "Other" having a text entry to be completed.

\noindent \textbf{Section 5 - Demographics:} A questionnaire was designed to gather participants' demographics information, as well as other information that could affect their attitude and behaviour during COVID-19, which included: (a) age, (b) gender, (c) if participants had pets/type of pets, (d) number of people in their household, (e) number of children in their household, (f) if they have a relative who is over 80 and lives alone (if yes, how can a social robot help them?), (g) if they have a relative who is over 80 and lives in a care home (if yes, how can a social robot help them?), (h) how can a social robot help them stay connected with an older relative, (i) how they feel about social isolation and COVID-19 (responses ranged from their life is completely and negatively affected to their life is completely and positively affected), and how they feel about the way their life has changed, (k) aside from work communications, how many people did they connect with and if this number was changed due to COVID-19, (l) whether they have ever moved to a new country and lived there for a significant period for a purpose other than a holiday (as an indication of their ability to adapt to significant changes in their lives), (m) how stressed/anxious they were due to COVID-19, (n) how much they followed social isolation rules, and (o) if they thought there were differences between social robots and conversational virtual agents, to explain which one they prefer, and to explain why.\footnote{Note that only a subset of these results, which were appropriate for the scope of this paper are discussed due to page limits.}

On each page of the questionnaire, attention and sanity checks were added. For example, we presented the same question with the opposite direction of the scale for the answers, or questions with clear answers, such as ``How much do you think that drinking water is liquid" or ``Do you agree that Tuesdays are considered weekends in North America?".

Finally, we intended for participants to complete the study on a positive note. Thus, with the purpose of trying to change participants' mood after thinking about the questions mentioned above, we asked them about their favourite animal, favourite cartoon character, and favourite movie. These questions were included at the very end and served solely as an attempt to end the study on a positive note.

\subsection{Procedure}

Upon reading the information letter and signing the consent form, participants were directed to the questionnaire. They completed the 5 sections of the questionnaire mentioned above. Afterwards they received the completion code and instructions on how to submit the HIT\footnote{Human Intelligence Task}.


\subsection{Participants}
The first study was conducted May-June 2020. 110 participants were recruited on Amazon Mechanical Turk. Participation was limited to Canada\footnote{Participation was limited to Canada, where rules on social distancing were precisely defined and followed by the majority of people.} and those who had an approval rate over 97\% (to minimize the risk of getting low quality responses) and had completed at least 100 HITs\footnote{A Human Intelligence Task (HIT) is a task on Mechanical Turk (MTurk), which is completed by the volunteers on MTurk.} (to ensure familiarity with the interface). Data of five participants were removed as they failed the attention checks. The data of three participants' were removed due to repeated participation. This finally resulted in 102 participants (41 Female, 61 Male; age: [18,66], mean: 34.6 yrs). The number of people in participants' household ranged from 1 to 8, with one not indicating a number and reporting ``more than 5". The number of children in participants' household ranged from 0 to 6, with one not indicating a number and reporting ``more than 5". The majority (68 participants) reported to have no children in their household.

The second study was conducted in January 2021. By checking the MTurk IDs prior to participation, we ensured that those who participated in the first study will not participate again. Recruitment criteria on Amazon Mechanical Turk was the same as before: 97\% approval rate based on at least 100 HITs. But as country was also limited as before, we were not able to recruit more than 70 participants for the second study (who did not participate in the first one); therefore, the criteria was changed to an approval rate of 95\% based on at least 50 HITs. A total of 138 participants were recruited (60 female, 77 male, and 1 unknown). Five participants failed the attention checks and their data were removed from the study. One participant's data was removed due to missing data. This left 132 participants (59 Female, 72 Male, and 1 unknown; age: [25,72], mean: 33.6 yrs). The reported number of people in participants' household ranged from 1 to 26. The number of children in participants' household ranged from 0 to 5. 

 Full Ethics clearance was received from the University of Waterloo's Research Ethics Committee prior to running the studies. Participants were notified about the nature of the questions under foreseeable risks in the consent form and were informed that the questions might make them think about different aspects of social isolation during COVID-19 and the feeling of loneliness. They were also given the option to skip questions (without it affecting the remuneration) or to stop at any point.

\section{Results}
First, we will discuss how participants' perception of social robots changed as a result of COVID-19, and how participants' loneliness level and this change in perception affected the tendency of purchasing a social robot. Afterwards, we will discuss the tasks that the participants preferred for a social companion robot and the elements/capabilities that they considered to be important for the robots, along with their preferred appearance for a social robot. Finally, we will present results on how participants' personality affected adapting to COVID-19 in general. We did not observe any effect of having a pet, or the number of people and children living in participants'  households on any of the measures, therefore these results are not discussed in this section.

\begin{figure}[t]
\centering
\begin{tabular}{cc}
     \includegraphics[width=0.5\linewidth]{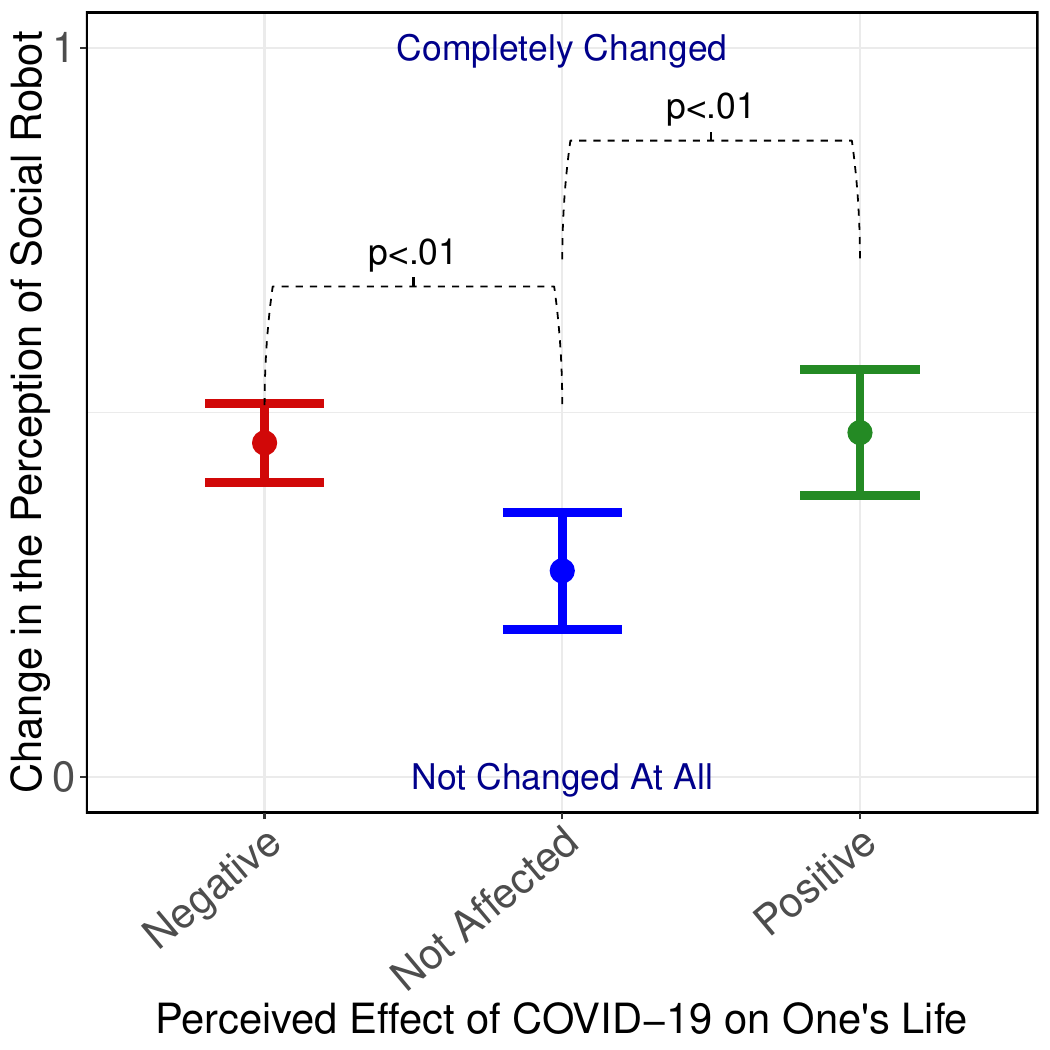}   & 
        \includegraphics[width=0.5\linewidth]{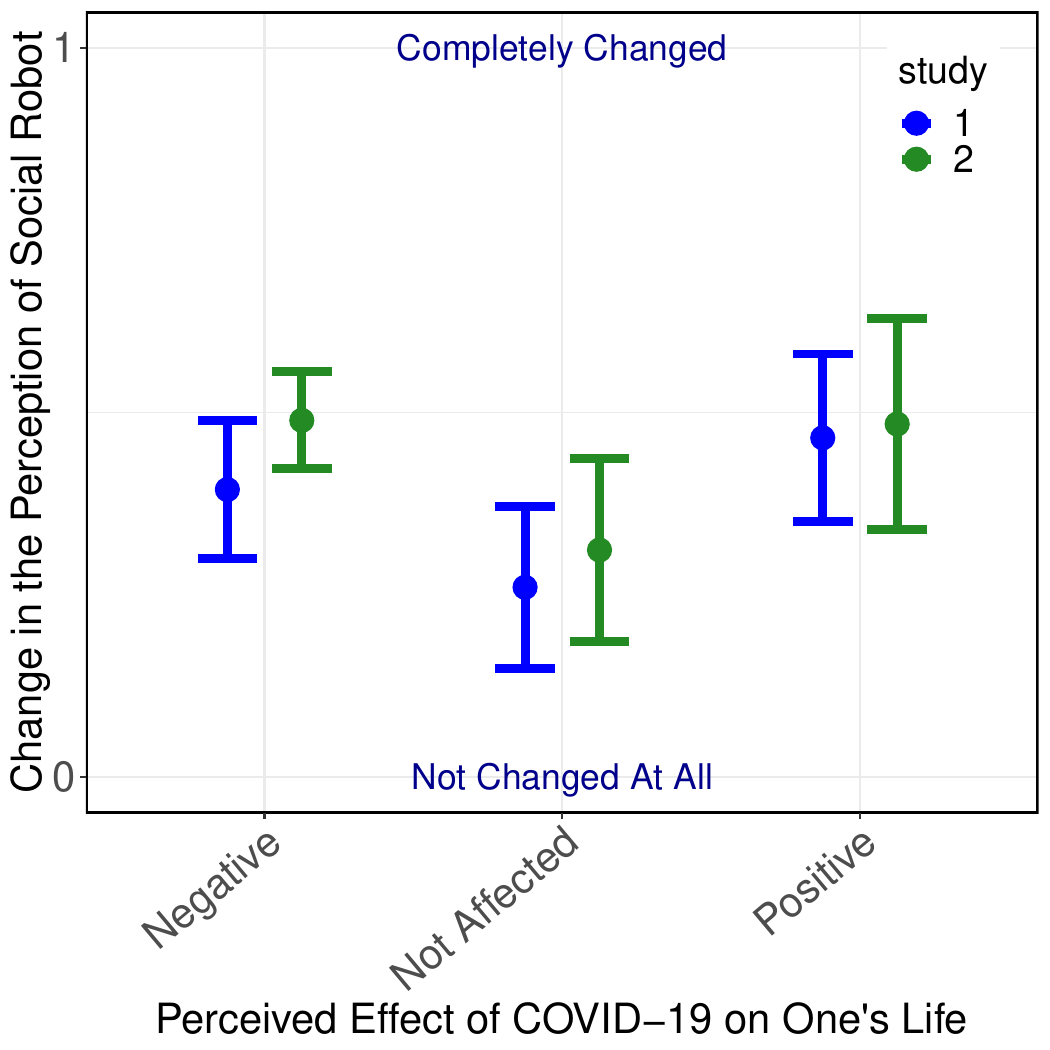}\\
  (a)   & (b) \\
\end{tabular}

    \caption{Change in the perception of social robots as a result of the reported change in people's lives due to COVID-19 for (a) both studies combined, and (b) each study. 95\% confidence intervals are visualized. Both a positive and a negative change significantly increased the change in perception of companion robots (see Table~\ref{tab:perceptionChange}). Note that in study 2 the number of people who reported a positive change decreased, therefore, the smaller sample in the positive group could have led to a larger confidence interval. In study 1 we had 44, 29, and 29 participants in ``Negative", ``Not affected", and ``Positive" groups, respectively. Whereas these numbers were 91, 23, and 18 in study 2.)}
    \label{fig:perceptionChange}
\end{figure}

\subsection{COVID-19 and Perception Change}

Experiencing a change in one's lifestyle as a result of the pandemic (whether positive or negative) led to a change in the perception of benefits of companion robots. This change was significantly higher than the change reported by those who indicated that their lives were not affected by COVID-19 (see Figure~\ref{fig:perceptionChange}). 
\begin{table}
\begin{center}
\small
\caption{\label{tab:perceptionChange} Linear Regression model predicting perception changed based on participants' lifestyle change due to COVID-19, and whether they had a relative over 80 who lives alone or at a care center. Age, gender, and Big 5 personality were controlled for but removed as they did not improve the model (neither had a significant effect on perception change).}
\renewcommand{\arraystretch}{1}
\begin{tabular}{lrrrr}
Covariate &  \multicolumn{1}{c}{Estimate} & \multicolumn{1}{c}{SE} & \multicolumn{1}{c}{t} & \multicolumn{1}{c}{Pr ($>|t|$)}\\ \hline
Intercept &  219.086  &  105.725  & 2.072 & $<.05$\\ 
Study     &    62.358  &   41.603  & 1.499 & 0.135 \\   
loneliness    &  -2.894  &  4.048 & -0.715 & 0.475 \\   
feelIsolationNegative &  158.686    & 50.591&   3.137& $<.01$\\ 
feelIsolationPositive  & 178.476  &   61.531 &   2.901 & $<.01$\\ 
OlderRelativeAloneTRUE  & 97.642  &  42.503 &  2.297 & $<.05$\\ 
\hline\end{tabular}
\end{center}
\end{table}

\begin{figure}
    \centering
    \begin{tabular}{cc}
        \includegraphics[width=0.5\linewidth]{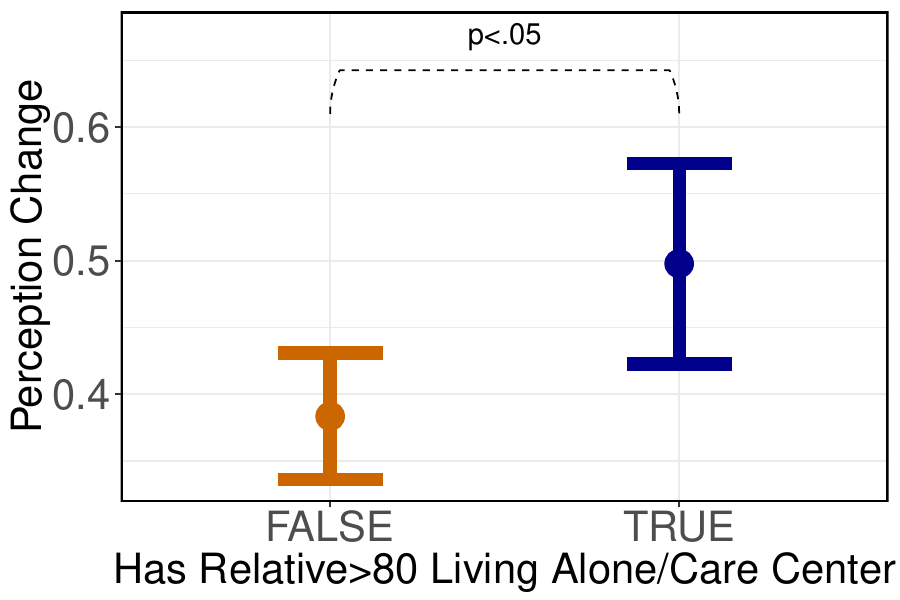}  &  \includegraphics[width=0.5\linewidth]{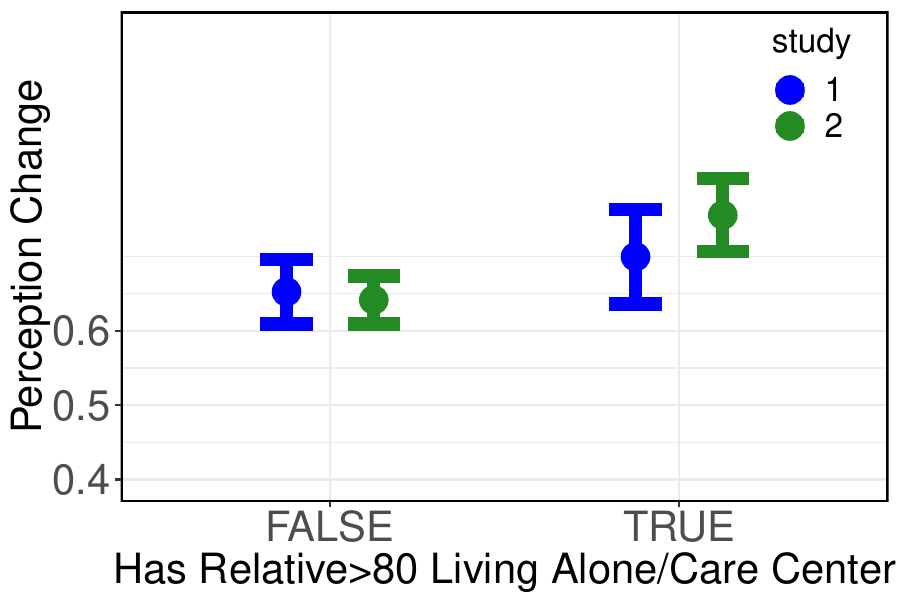} \\
         (a) & (b) \\
    \end{tabular}
      
    \caption{Reported perception change based on whether the participants had an older relative who either lived alone or at a care center (a) for the combined data, and (b) for each study.}
    \label{fig:olderRelative}
\end{figure}

To study the significance of this difference, a linear model was fit to predict perception change, based on the change in participants' lifestyle, whether they had an older relative (over 80) who lived alone or at a care home, and study (i.e., time of the study). Table~\ref{tab:perceptionChange} shows the results. Both a positive ($t=3.016, p<.01$) and a negative change ($t=3.228, p<.01$) significantly affected the perception change (confirming H1). Further, having a relative who either lives alone or at a care center significantly and positively affected the perception change ($t=2.322, p<.05$; confirming H2; also see Figure~\ref{fig:olderRelative}). We did not see an effect of study (i.e., time of study) on perception change (supporting H6).

\begin{figure}[t]
    \begin{tabular}{cc}
          \includegraphics[width=0.47\linewidth]{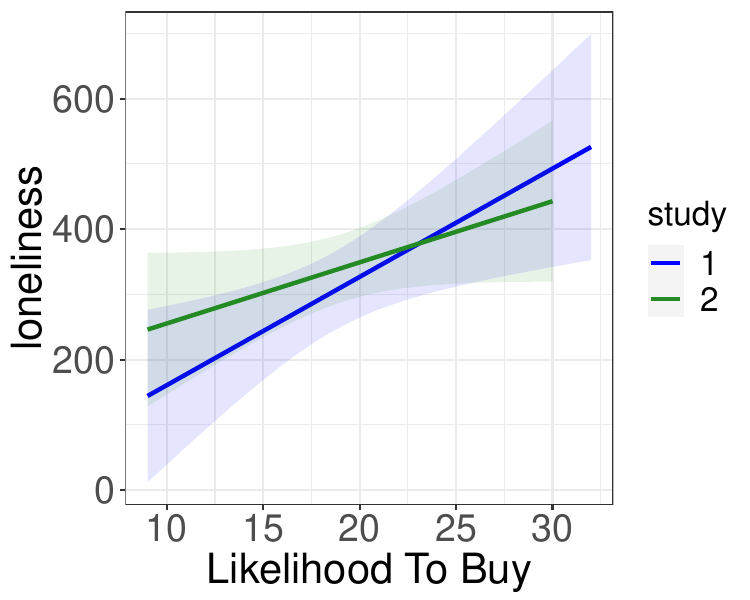} &  
            \includegraphics[width=0.47\linewidth]{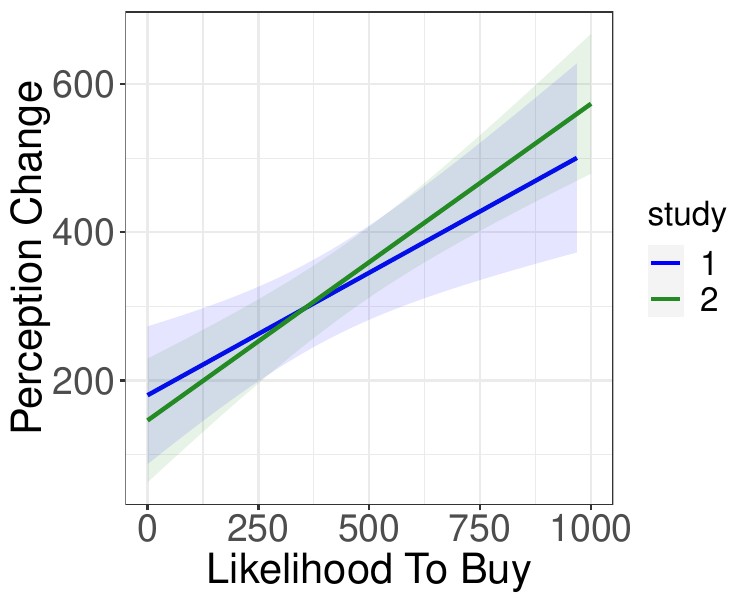}\\
     (a)    &  (b)
    \end{tabular}

\caption{Reported likelihood to purchase a social robot based on (a) the measured level of loneliness and (b) change in perception of social robots due to COVID-19 (see Table\ref{tab:loneliness} for more details). }
    \label{fig:loneliness}
\end{figure}

\begin{table}
\begin{center}
\caption{\label{tab:loneliness} Linear Regression model predicting the likelihood to purchase a social robot based on the loneliness level, study (time of the study), Big 5 personality traits, age, gender, and participants' perception change regarding social robots due to COVID-19. Loneliness and the change in perception were not correlated.}
\renewcommand{\arraystretch}{1}
\begin{tabular}{lrrrc}
Covariate &  \multicolumn{1}{c}{Estimate} & \multicolumn{1}{c}{SE} & \multicolumn{1}{c}{t} & \multicolumn{1}{c}{Pr ($>|t|$)}\\ \hline
Intercept   &    -135.271 &  350.043 & -0.386 & .700\\    
Study          &      6.034 &   36.810 &   0.164 & .870    \\
Loneliness      &    15.570&    4.368 &   3.564 & $<.001$\\
Extraversion     &   11.906&   15.266 &   0.780 & .436    \\
Agreeableness     &  24.378 &   16.952 &   1.438 &.152  \\  
Conscientiousness  & 23.178 &   17.046 &   1.360 &.175\\    
Emotionality    &     0.447 &   17.600 &   0.025 &.980 \\   
Openness        &   -54.713 &   17.878 &  -3.060 &$<.01$\\
PerceptionChanged &   0.395 &    0.057 &   6.883 &$<.0001$\\
Gender:Male &   90.861 &   39.554&   2.297 &$<.05$\\
Gender:DidNotShare &   23.860&  277.009 &   0.086& .931\\ 
Age             &    -1.661 &    1.629 &  -1.020 & .309\\   

\hline\end{tabular}
\end{center}
\end{table}
This change in perception of the benefits of social robots due to COVID-19 also significantly affected the reported likelihood of purchasing a social robot ($t=6.883, p<.0001$; confirming H3; see Table~\ref{tab:loneliness}). Furthermore, loneliness, along with this perception change, significantly and positively affected the likelihood of purchasing a companion robot ($t=3.564, p<.001$
). Results are shown in Table~\ref{tab:loneliness} and Figure~\ref{fig:loneliness}. This confirmed H4. This likelihood was also affected by one of the Big~5 personality traits: openness ($t=-3.060, p<.01$), but unlike what was expected, as openness increased, the likelihood to purchase a social robot decreased. We also noticed gender differences as male participants reported a significantly higher likelihood to purchase a social robot as compared with female participants ($t=2.297, p<.05$). Here as well, we did not see an effect of study (i.e., time of study) on perception change (supporting H6).

\subsection{Preferred Tasks for a Companion Robot}

\begin{figure}
    \centering
       \includegraphics[width=0.7\linewidth]{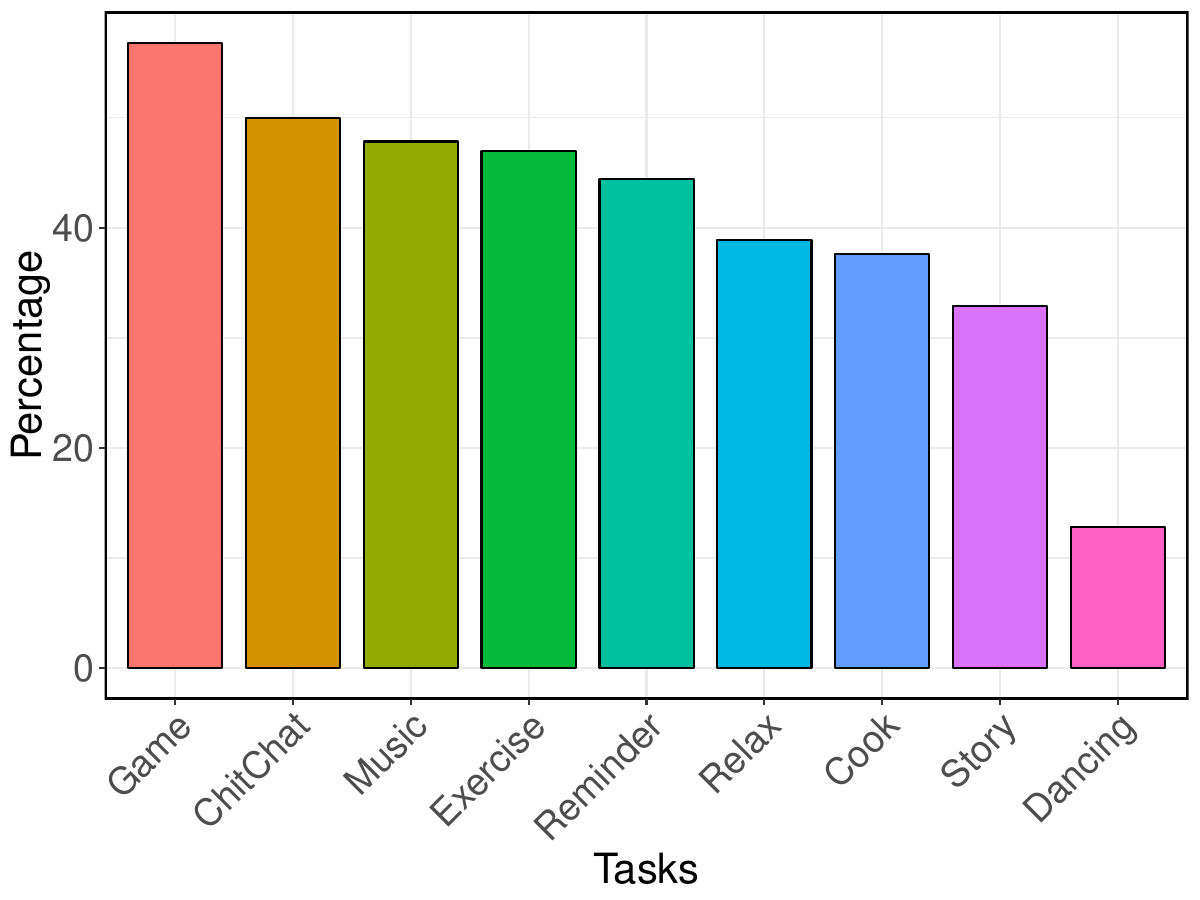}
    \caption{Tasks that participants preferred to use a companion robot for. Percentage shows the percentage of participants who selected the task. }
    \label{fig:tasks}
\end{figure}

Accepting social robots as companions might also depend on the tasks that they are designed for. Therefore we studied participants' responses related to the preferred task for a companion robot. Figure~\ref{fig:tasks} shows the results. While we did not find any strong preference towards a specific task, playing games, getting involved in a chitchat, playing music, helping in exercises, and helping with reminders were considered as the most preferred tasks for a companion robot, while ``dancing" was the least preferred. 

We performed binomial tests to study whether the differences were significant. Dancing was selected significantly less than all other options ($p<.0001$), and getting involved in games was selected significantly more than helping with reminders ($p<.05$), relaxation ($p<.01$), cooking ($p<.01$), story telling ($p<.0001$), and dancing ($p<.0001$).

In addition, 18 participants selected other, either alone or along with other choices, four of whom  indicated that they were not interested in using a companion robot and many others did not provide any explanation. The suggestions for ``other" included cleaning the house, helping with household choirs, scheduling appointments, providing information about the weather and news, and giving random factoids.

\subsection{Important Elements for a Companion Robot}

\begin{figure}
\centering
     \includegraphics[width=0.6\linewidth]{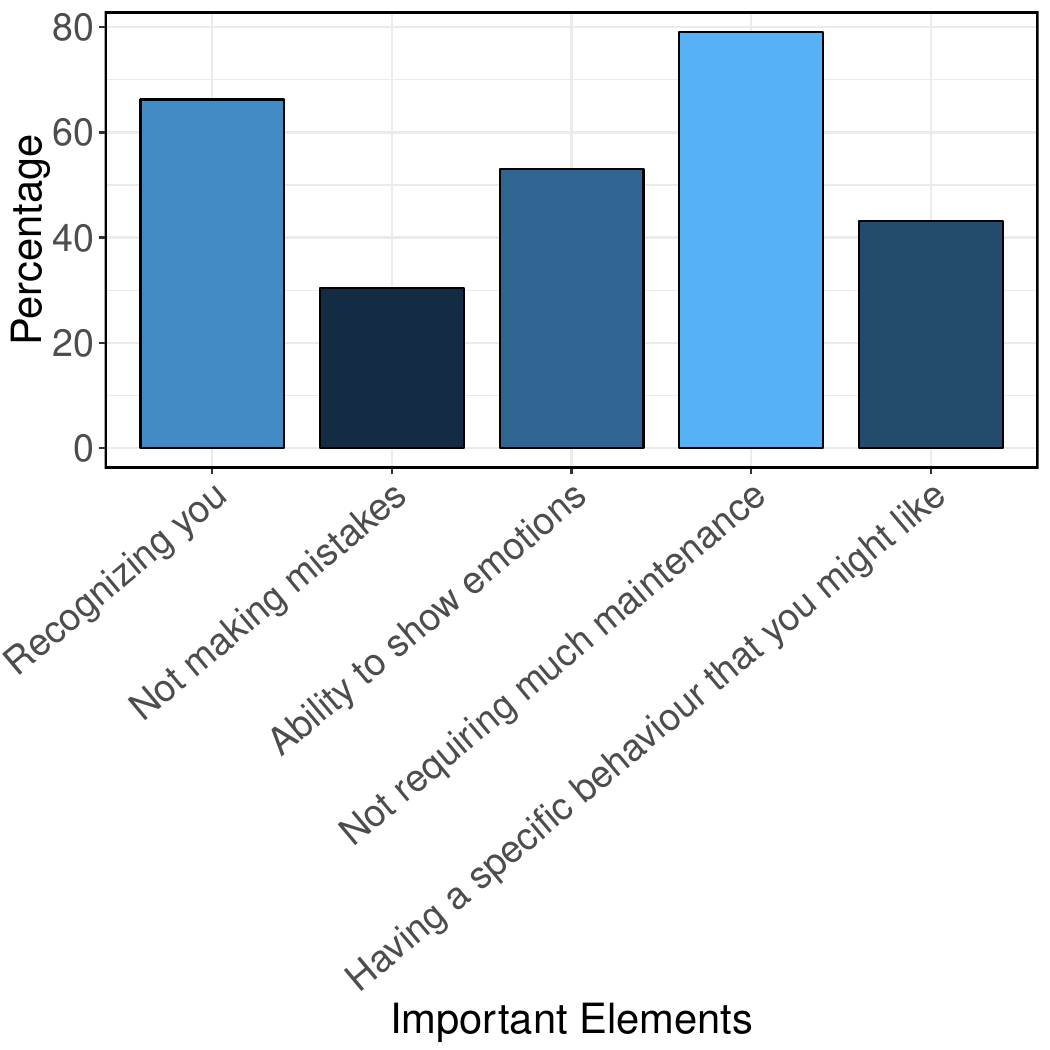}
    \caption{Important elements of a companion robot as indicated by the participants. Percentage shows the percentage of participants who selected the element.}
    \label{fig:elements}
\end{figure}

We further studied different social and technical capabilities of  robots and asked which one/ones the participants believed to be the most important in a companion social robot. We found ``not requiring any maintenance" and ``recognizing you" to be two of the important elements for social robots, and, interestingly, social capabilities such as ``recognizing you", ``ability to show emotions", and ``having a specific behaviour that one prefers" to be considered even more important than technical soundness, i.e., ``not making mistakes". Thus, it indicates that participants prefer a social companion robot that is technically robust but they do not necessarily ask for an advanced level of rational intelligent behaviour. The results are shown in Figure~\ref{fig:elements}.

We further conducted binomial tests to check whether the differences were significant. Among these elements, ``not requiring much maintenance" and ``recognizing you" were selected significantly more than all others (significance levels ranged from $p<.05$ to $p<.0001$; the difference between these two elements were not significant). Also, ``Not making mistakes" was selected significantly less than all other elements (significance levels changed from $p<.05$ to $p<.0001$). This suggested that along with not requiring maintenance, social capabilities might be even more important than not making mistakes in a social companion robot. This confirmed H5.

In addition to these options, 22 participants selected ``Other" and entered the elements that were important to them. Three of the inputs were about the application areas (i.e., playing guitar, fetching things, and doing chores/cooking), which reflects a misunderstanding of this question. Two others indicated that they were not interested in a robot regardless, so there were not any element that was important for them. One did not explain the choice of other and did not select any other element. The remaining others suggested additional elements. Two of them asked for the robot to be dynamic and helpful (which would also refer to a robot's functionality). Two others pointed out maintenance and safety as key elements: ``safe around home and pets" and
``easy to assemble, charge, and repair if needed". Seven addressed the communication capabilities of the robot (i.e., ``natural conversation", ``being able to communicate freely (not experience the same comprehension problems as with Alexa)", "doesn't not understand us" [sic], ``sounding natural not robotic", ``the ability to converse", "being able to react to a specific situation with some Randomnise" [sic], and ``It is absolutely imperative that the ilusion of talking to something sentient and not just a bunch of lines of code be maintained, otherwise it is a tool, not a companion." [sic]). Furthermore, four participants emphasized the importance of other social capabilities of a companion robot (i.e., ``Mimicking empathy and giving consolation", ``a kind \textit{personality}", ``understand my mood", and
``remembering me, getting to know me, building on a relationship") as key elements for a companion robot. Lastly, one participant indicated that ``selecting gender" would be an important element.

\subsection{Preferred Appearance of a Companion Robot}

We also studied participants' appearance preferences for a companion robot. Both a human-like and an animal-like appearance were similarly preferred (110 versus 92 votes). Thirty two participants selected ``Other" and proposed their own preferred appearance. 17 participants indicated robot-like or machine-like, with some giving more information (e.g., ``A retro styled robot from the 1950s" and ``robot like ie tv / movie like in appearance" [sic]). Further, one response indicated animal-like ("Anthropomorphic fox character") and two indicated that they would prefer no robot. Others suggestions included (a) any shape but human-like, (b) not similar to anything living, (c) in the form of a speaker or a box, (d) object-like, like a box, ball, or disk, (e) not like a life form.  e.g., like a tennis ball or hockey puck, (f) the form should be dictated by the function of the robot, (g) the shape would not be important, and (h) not human but should have eyes.

\subsection{Personality Traits and Adapting to COVID-19}

Aside from the effects of the personality traits on perception of companion robots, Big 5 traits affected participants' general attitude towards the pandemic. Scoring high on the extroversion trait significantly increased being stressed as a result of COVID-19 ($t=3.192,  p<.01$), as reported by the participants, while scoring high on emotional stability significantly decreased participants' stress ($t=-5.800, p<.0001$). Also, those who reported to follow the rules better reported to have less stress about COVID-19 ($t=4.147, p<.0001$). While not significant, the stress levels seemed to be higher in Study 2, as compared with the first study ($t=1.909, p=.057$). 

Furthermore, extroversion significantly and negatively affected following the social isolation rules ($t=-2.539,  p<.05$), while agreeableness significantly increased following the rules ($t=3.848, p<.001$). Also, conscientiousness did not affect how well people follow the social isolation rules ($t=-0.053, p=.957$). Finally, the results were different across two studies and participants reported to follow rules significantly more in study~2 as compared with study~1 ($t=2.711, p<.01$).~\footnote{p-values were calculated through two linear models predicting stress and how much one follows rules based on participants' personality traits}

We also observed similar effects of Big 5 traits on participants' loneliness levels as was reported by Buecker et el.~\cite{buecker2020loneliness}: extroversion, agreeableness, and emotional stability significantly decreased loneliness. We did not find an effect of gender or age on loneliness. Further, we did not find an effect of study (i.e., when the study was conducted) on the loneliness levels and conscientiousness did not affect loneliness ($t=-0.072, p=.943$ ). Table~\ref{tab:lonelinessPersonality} shows these results.

\begin{table}
\begin{center}
\caption{\label{tab:lonelinessPersonality} Linear Regression model predicting loneliness based on participants' personality traits and the experience of immigration.}
\renewcommand{\arraystretch}{1}
\begin{tabular}{lrrrl}
Covariate &  \multicolumn{1}{c}{Estimate} & \multicolumn{1}{c}{SE} & \multicolumn{1}{c}{t} & \multicolumn{1}{c}{Pr ($>|t|$)}\\ \hline
Study2         &    0.091 &   0.563 &   0.161 & 0.872 \\
FollowRules     &  -0.001 &   0.002 &  -0.925&  0.356 \\
Immigration     &    0.899 &   0.585 &   1.536&  0.126 \\ 
Extraversion     & -0.829 &   0.224 &  -3.697 & $<.001$\\
Emotionality     & -1.315 &   0.234 &  -5.630 &$<.0001$\\
Conscientiousness &-0.018 &   0.255 &  -0.072 &0.943 \\   
Openness          &-0.241 &   0.268 &  -0.899 &0.370 \\    
Agreeableness     &-0.635&    0.257 &  -2.468 & $<.05$\\
\hline\end{tabular}
\end{center}
\end{table}

\section{Discussion}

\begin{itemize}
   \item[] \textit{``I think my perception has changed because I have never fully understood what it feels like to be isolated for such a long period. When thinking about social isolation I usually thought of people in remote areas or people who were in care homes without many visitors. Now it is easier to understand what people living alone must feel like and why having some sort of social interaction is important."\footnote{Examples of participants' quotes related to their perception change.}}
    \item[]
    \item[] \textit{``The COVID-19 pandemic has revealed to me that certain individuals are deeply distressed by isolation and the inability to socialize with others regularly or in the conventional way. If there is a tool available that can help others feel less isolated, I'm all for it."}
     \item[] 
    \item[] \textit{``It was changed because recently it has been difficult to meet up with other people and socialize. This has put a strain on people's mental health, including mine. For this reason, I feel that a social robot would be an appropriate way to spend time. At this point, I am okay with spending time with anything that will keep me busy and interested."}       
\end{itemize}

There has been an ongoing debate regarding the benefits of using companion robots. While social robots are shown to be effective in reducing loneliness, concerns such as using companion robots may lead to neglect from family members, reduce human interactions, or replace humans and lead to loss of jobs have negatively affected society's attitude towards companion robots~\cite{wang2017robots,gnambs2019robots}. 

In this paper, we presented two studies conducted 8 months apart and hypothesized that COVID-19 could have changed society's perception of social robots, as during the self-isolation period people are more likely to experience social isolation and think about the consequences of social isolation, either due to the change they experienced themselves (RQ1), or thinking about others, such as older relatives who live alone or at a care center (RQ2). As a result, they may be able to see social robots as additions and complements, as opposed to replacements for human contact, which can lead to a more positive attitude towards social robots, and ultimately a higher acceptance of social companion robots in the future. Also, with loneliness being one of the situations that companion robots could help with, we asked if loneliness can increase one's tendency to purchase a social robot (RQ3). 

Furthermore, to understand general preferences related to companion robots, we asked whether people have specific preferences towards the tasks designed for companion robots (RQ4), and if social elements of companion robots (e.g., the ability to show emotions) are important to them, and how this importance compares to other elements such as a robot's technical accuracy (RQ5). Finally, as it can be informative for designing social companion robots during pandemics, we asked whether people's personality (Big 5) can affect their behaviour during a pandemic and their attitude towards social robots (RQ6).

Two studies were conducted to ensure that the results were robust and did not depend on participants' level of experience with the pandemic (RQ7), as we hypothesized that the effects found will be long term.

The second study replicated results of the first study (H6 confirmed). Our results provided evidence about a change of perception of the benefits of social companion robots, and suggested that any change experienced during COVID-19 (negative or positive) affected perception of social companion robots (H1 confirmed), as well as the reported likelihood of buying a companion robot (H3 confirmed). While it was impossible to measure perception retrospectively, before and during the pandemic to compare the differences, we relied on participants' self-reported perception change. This change was in fact significantly different between those whose lives were not affected and those whose lives were affected by COVID-19.

We also noticed that those who had an older relative that either lived alone or at a care center had a significantly higher perception change (H2 confirmed). Similar to experiencing a change in the lifestyle, having an older relative who lives alone or is at a care center during the pandemic~\footnote{Note that visiting care centers by family members was prohibited during the time this study was conducted in Canada.} can emphasize the benefits of social robots during situations where social contact is impossible, and that might explain this change on the attitude towards companion robots. Thus, experiencing a pandemic like COVID-19 could have positively changed society's attitude towards social companion robots, by emphasizing their benefits in a situation where human contact was clearly not possible. It is important to note that while our participants' ages ranged from 19 to 72 years, supporting that this perception change happened in both younger and older adults, future studies are needed to understand how this change would affect purchase of social robots for an older relative, and whether older relatives would be willing to accept and use social robots in these situations.

Other than this perception change, loneliness was another important factor that also increased the likelihood of buying a social companion robot (H4 confirmed). This is important as it can help adoption of social robots among those who have a higher level of loneliness and would benefit the most from the presence of a social companion robot. Loneliness was in fact previously reported to affect the perceived social presence of social agents, after people interacted with an agent~\cite{lee2006physically}: loneliness led to feeling a higher social presence of the social agents and led to a more positive social response from the participants~\cite{lee2006physically}, which suggests that the benefits of social robots are experienced more by those who experience loneliness, who also, based on our study's findings, were more likely to adopt a social robot. Interestingly, openness  trait of Big~5 personality had a negative effect on the tendency to purchase a social robot. This was unexpected as openness can also reflect one's openness to new experiences, and it is expected to positively affect the tendency to purchase a social robot. Future work is needed to understand why this effect was observed.

Our study also pointed out interesting and general findings on how people with different personalities adopted to COVID-19. Although we did not observe any effect of personality traits on participants' change of perception of social robots, personality traits affected participants' purchase tendency of social robots, stress levels, how much they followed rules in general, and their loneliness levels. All of these findings can be informative for the behaviour and functionality of a social companion robot during a pandemic. For example, these results suggest that extroverts could benefit highly from the presence of a companion robot during pandemics, especially one that can help reduce stress levels (e.g., by providing therapeutic support), and which could also possibly educate them and encourage them to follow social rules. Further, those with a lower emotional stability, extroversion, and agreeableness could benefit from the presence of a social robot that can reduce their loneliness.

Lastly, we provided a summary of participants' preferences towards tasks for a companion robot, its appearance, and the important elements/capabilities for such a robot. The results pointed out that a variety of tasks are preferred in general for a social companion robot, with dancing being the least preferred, and games being the most preferred. Among the important elements, ``not making mistakes" was considered to be least important, and significantly less important as compared with some social aspects, such as recognizing the user and the robot's ability to show emotions (H5 confirmed). Further, maintenance and the ability to recognize one seemed to be the most important elements, selected significantly more than all other elements of a social robot. Regarding appearance, both animal-like and human-like appearances were equally preferred; a choice which we believe needs to be made based on the concrete tasks designed for a social robot and needs to be studied in the future work.

It is important to emphasize that it is not our intention to promote the replacement of human contact with social robots, but as the current pandemic is showing, social robots might play an important part in situations where direct human contact is prohibited (e.g., during pandemics) or impossible (e.g., for socially isolated older adults). While it was not the focus of our study, in addition to providing social communication and assistance, robots could even provide tactile contact for therapeutic purposes, which (while still mechanical in nature and lacking the nature of touching a biological, sentient being) has been shown to be beneficial~\cite{Paro}.





\section{Limitations and Future Work}

Our studies had several limitations. First, we did not have data on people's attitude before and after COVID-19, we could only measure the change self-reported by the participants. Similarly, as the studies were conducted during the pandemic and during the period of self-isolation, we relied on participants' responses, and only loneliness and personality traits were measured through standard questionnaires (as opposed to direct questions). Furthermore, due to the strict rules on social distancing which were followed by the majority of people in Canada, participation was limited to Canada. Finally, our selection of options for appearance, tasks, and other robot elements were based on the existing literature, but more of those tasks and features could be explored. Also, in all of our multiple-choice questions we provided an option where the participants could input their preference beyond the existing selections. However, it is still possible that these results were affected by the provided choices and adding more options may change these results.

Future studies would be beneficial to understand how COVID-19 affected perception of social robots in other countries, and what can be learned from the perception change due to this pandemic to positively influence perception of social robots in other contexts and situations.

\section{Conclusion}

While social companion robots have been shown to have many benefits and can improve health and well-being, society's attitude towards them has not always been positive. In this study we asked how COVID-19 affected society's perception of social robot, as the pandemic was an example of a situation where social contact among humans could be impossible, and might have emphasized the benefits of using social companion robots as a ``complement", as opposed to a  ``replacement" (which is commonly held believe and is a strong reason for negative attitudes  towards social robots). The results of two studies conducted within 8 months of each other during the pandemic suggested that a change in one's life due to COVID-19 has changed the (self-reported) attitude towards social robots, as well as the tendency to purchase a social robot. These findings are promising, as a positive change in perception of social robots can increase their adoption, which would be especially advantageous for older adults who live alone or other socially isolated individuals, and can improve health and well-being. 

\section*{Acknowledgements}
 This research was undertaken, in part, thanks to funding from the Canada 150 Research Chairs Program and the Network for Aging Research at the University of Waterloo. We thank Sami Alperen Akgun for his help with the implementation of the questionnaire.

%
%
%
\bibliographystyle{splncs04}
\bibliography{references}

\begin{thebibliography}{10}
\providecommand{\url}[1]{\texttt{#1}}
\providecommand{\urlprefix}{URL }
\providecommand{\doi}[1]{https://doi.org/#1}

\bibitem{abdollahi2017pilot}
Abdollahi, H., Mollahosseini, A., Lane, J.T., Mahoor, M.H.: A pilot study on
  using an intelligent life-like robot as a companion for elderly individuals
  with dementia and depression. In: 2017 IEEE-RAS 17th International Conference
  on Humanoid Robotics (Humanoids). pp. 541--546. IEEE (2017)

\bibitem{alpass2003loneliness}
Alpass, F.M., Neville, S.: Loneliness, health and depression in older males.
  Aging \& mental health  \textbf{7}(3),  212--216 (2003)

\bibitem{armitage2020covid}
Armitage, R., Nellums, L.B.: Covid-19 and the consequences of isolating the
  elderly. The Lancet Public Health  \textbf{5}(5), ~e256 (2020)

\bibitem{baisch2017acceptance}
Baisch, S., Kolling, T., Schall, A., R{\"u}hl, S., Selic, S., Kim, Z.,
  Rossberg, H., Klein, B., Pantel, J., Oswald, F., et~al.: Acceptance of social
  robots by elder people: does psychosocial functioning matter? International
  Journal of Social Robotics  \textbf{9}(2),  293--307 (2017)

\bibitem{bennett2017robot}
Bennett, C.C., Sabanovic, S., Piatt, J.A., Nagata, S., Eldridge, L., Randall,
  N.: A robot a day keeps the blues away. In: 2017 IEEE International
  Conference on Healthcare Informatics (ICHI). pp. 536--540. IEEE (2017)

\bibitem{bishop2019social}
Bishop, L., van Maris, A., Dogramadzi, S., Zook, N.: Social robots: The
  influence of human and robot characteristics on acceptance. Paladyn, Journal
  of Behavioral Robotics  \textbf{10}(1),  346--358 (2019)

\bibitem{broadbent2009acceptance}
Broadbent, E., Stafford, R., MacDonald, B.: Acceptance of healthcare robots for
  the older population: Review and future directions. International journal of
  social robotics  \textbf{1}(4), ~319 (2009)

\bibitem{broadbent2012attitudes}
Broadbent, E., Tamagawa, R., Patience, A., Knock, B., Kerse, N., Day, K.,
  MacDonald, B.A.: Attitudes towards health-care robots in a retirement
  village. Australasian journal on ageing  \textbf{31}(2),  115--120 (2012)

\bibitem{buecker2020loneliness}
Buecker, S., Maes, M., Denissen, J.J., Luhmann, M.: Loneliness and the big five
  personality traits: A meta-analysis. European Journal of Personality
  \textbf{34}(1),  8--28 (2020)

\bibitem{chandra2019children}
Chandra, S., Dillenbourg, P., Paiva, A.: Children teach handwriting to a social
  robot with different learning competencies. International Journal of Social
  Robotics pp. 1--28 (2019)

\bibitem{dautenhahn2003roles}
Dautenhahn, K.: Roles and functions of robots in human society-implications
  from research in autism therapy. Robotica  (2003)

\bibitem{de2017they}
De~Graaf, M., Allouch, S.B., Van~Diik, J.: Why do they refuse to use my robot?:
  Reasons for non-use derived from a long-term home study. In: 2017 12th
  ACM/IEEE International Conference on Human-Robot Interaction (HRI. pp.
  224--233. IEEE (2017)

\bibitem{derek2012socially}
Derek, M., Chan, J., Nejat, G.: A socially assistive robot for meal-time
  cognitive interventions. Journal of Medical Devices  \textbf{6}(1) (2012)

\bibitem{douglas2020mitigating}
Douglas, M., Katikireddi, S.V., Taulbut, M., McKee, M., McCartney, G.:
  Mitigating the wider health effects of covid-19 pandemic response. Bmj
  \textbf{369} (2020)

\bibitem{gerst2015loneliness}
Gerst-Emerson, K., Jayawardhana, J.: Loneliness as a public health issue: the
  impact of loneliness on health care utilization among older adults. American
  journal of public health  \textbf{105}(5),  1013--1019 (2015)

\bibitem{gnambs2019robots}
Gnambs, T., Appel, M.: Are robots becoming unpopular? changes in attitudes
  towards autonomous robotic systems in europe. Computers in Human Behavior
  \textbf{93},  53--61 (2019)

\bibitem{goh2020vision}
Goh, P.S., Sandars, J.: A vision of the use of technology in medical education
  after the covid-19 pandemic. MedEdPublish  \textbf{9} (2020)

\bibitem{gosling2003very}
Gosling, S.D., Rentfrow, P.J., Swann~Jr, W.B.: A very brief measure of the
  big-five personality domains. Journal of Research in personality
  \textbf{37}(6),  504--528 (2003)

\bibitem{de2016long}
de~Graaf, M.M., Allouch, S.B., van Dijk, J.: Long-term acceptance of social
  robots in domestic environments: Insights from a user's perspective. In: AAAI
  Spring Symposia (2016)

\bibitem{hays1987short}
Hays, R.D., DiMatteo, M.R.: A short-form measure of loneliness. Journal of
  personality assessment  \textbf{51}(1),  69--81 (1987)

\bibitem{heerink2008influence}
Heerink, M., Kr{\"o}se, B., Evers, V., Wielinga, B.: The influence of social
  presence on acceptance of a companion robot by older people  (2008)

\bibitem{johnson2014measuring}
Johnson, J.A.: Measuring thirty facets of the five factor model with a 120-item
  public domain inventory: Development of the ipip-neo-120. Journal of Research
  in Personality  \textbf{51},  78--89 (2014)

\bibitem{kim2020s}
Kim, J.S., Kwak, S.S., Kang, D., Choi, J.: What's in a name? effects of
  category labels on the consumers' acceptance of robotic products. In:
  Proceedings of the 2020 ACM/IEEE International Conference on Human-Robot
  Interaction. pp. 599--607 (2020)

\bibitem{lee2006physically}
Lee, K.M., Jung, Y., Kim, J., Kim, S.R.: Are physically embodied social agents
  better than disembodied social agents?: The effects of physical embodiment,
  tactile interaction, and people's loneliness in human--robot interaction.
  International journal of human-computer studies  \textbf{64}(10),  962--973
  (2006)

\bibitem{li2020does}
Li, S., Xu, L., Yu, F., Peng, K.: Does trait loneliness predict rejection of
  social robots? the role of reduced attributions of unique humanness
  (exploring the effect of trait loneliness on anthropomorphism and acceptance
  of social robots). In: Proceedings of the 2020 ACM/IEEE International
  Conference on Human-Robot Interaction. pp. 271--280 (2020)

\bibitem{mannion2019introducing}
Mannion, A., Summerville, S., Barrett, E., Burke, M., Santorelli, A., Kruschke,
  C., Felzmann, H., Kovacic, T., Murphy, K., Casey, D., et~al.: Introducing the
  social robot mario to people living with dementia in long term residential
  care: Reflections. International Journal of Social Robotics pp. 1--13 (2019)

\bibitem{manyika2017jobs}
Manyika, J., Lund, S., Chui, M., Bughin, J., Woetzel, J., Batra, P., Ko, R.,
  Sanghvi, S.: Jobs lost, jobs gained: Workforce transitions in a time of
  automation. McKinsey Global Institute  \textbf{150} (2017)

\bibitem{moyle2014connecting}
Moyle, W., Jones, C., Cooke, M., O'Dwyer, S., Sung, B., Drummond, S.:
  Connecting the person with dementia and family: a feasibility study of a
  telepresence robot. BMC geriatrics  \textbf{14}(1), ~7 (2014)

\bibitem{nanevasystematic}
Naneva, S., Gou, M.S., Webb, T.L., Prescott, T.J.: A systematic review of
  attitudes, anxiety, acceptance, and trust towards social robots

\bibitem{nomura2006experimental}
Nomura, T., Kanda, T., Suzuki, T.: Experimental investigation into influence of
  negative attitudes toward robots on human--robot interaction. Ai \& Society
  \textbf{20}(2),  138--150 (2006)

\bibitem{nomura2009influences}
Nomura, T., Yamada, S., Kanda, T., Suzuki, T., Kato, K.: Influences of concerns
  toward emotional interaction into social acceptability of robots. In: 2009
  4th ACM/IEEE International Conference on Human-Robot Interaction (HRI). pp.
  231--232. IEEE (2009)

\bibitem{odetti2007preliminary}
Odetti, L., Anerdi, G., Barbieri, M.P., Mazzei, D., Rizza, E., Dario, P.,
  Rodriguez, G., Micera, S.: Preliminary experiments on the acceptability of
  animaloid companion robots by older people with early dementia. In: 2007 29th
  Annual International Conference of the IEEE Engineering in Medicine and
  Biology Society. pp. 1816--1819. IEEE (2007)

\bibitem{paetzel2020persistence}
Paetzel, M., Perugia, G., Castellano, G.: The persistence of first impressions:
  The effect of repeated interactions on the perception of a social robot. In:
  Proceedings of the 2020 ACM/IEEE International Conference on Human-Robot
  Interaction. pp. 73--82 (2020)

\bibitem{perugia2017modelling}
Perugia, G., Doladeras, M.D., Mallofr{\'e}, A.C., Rauterberg, M., Barakova, E.:
  Modelling engagement in dementia through behaviour. contribution for socially
  interactive robotics. In: 2017 International Conference on Rehabilitation
  Robotics (ICORR). pp. 1112--1117. IEEE (2017)

\bibitem{picking2017exploring}
Picking, R., Pike, J.: Exploring the effects of interaction with a robot cat
  for dementia sufferers and their carers. In: 2017 Internet Technologies and
  Applications (ITA). pp. 209--210. IEEE (2017)

\bibitem{poscia2018interventions}
Poscia, A., Stojanovic, J., La~Milia, D.I., Duplaga, M., Grysztar, M., Moscato,
  U., Onder, G., Collamati, A., Ricciardi, W., Magnavita, N.: Interventions
  targeting loneliness and social isolation among the older people: An update
  systematic review. Experimental gerontology  \textbf{102},  133--144 (2018)

\bibitem{van2019social}
Van~der Putte, D., Boumans, R., Neerincx, M., Rikkert, M.O., de~Mul, M.: A
  social robot for autonomous health data acquisition among hospitalized
  patients: an exploratory field study. In: 2019 14th ACM/IEEE International
  Conference on Human-Robot Interaction (HRI). pp. 658--659. IEEE (2019)

\bibitem{robb2020robots}
Robb, D.A., Ahmad, M.I., Tiseo, C., Aracri, S., McConnell, A.C., Page, V.,
  Dondrup, C., Chiyah~Garcia, F.J., Nguyen, H.N., Pairet, {\`E}., et~al.:
  Robots in the danger zone: Exploring public perception through engagement.
  In: Proceedings of the 2020 ACM/IEEE International Conference on Human-Robot
  Interaction. pp. 93--102 (2020)

\bibitem{robins2009isolation}
Robins, B., Dautenhahn, K., Dickerson, P.: From isolation to communication: a
  case study evaluation of robot assisted play for children with autism with a
  minimally expressive humanoid robot. In: 2009 Second International
  Conferences on Advances in Computer-Human Interactions. pp. 205--211. IEEE
  (2009)

\bibitem{robins2005robotic}
Robins, B., Dautenhahn, K., Te~Boekhorst, R., Billard, A.: Robotic assistants
  in therapy and education of children with autism: can a small humanoid robot
  help encourage social interaction skills? Universal access in the information
  society  \textbf{4}(2),  105--120 (2005)

\bibitem{vsabanovic2013paro}
{\v{S}}abanovi{\'c}, S., Bennett, C.C., Chang, W.L., Huber, L.: Paro robot
  affects diverse interaction modalities in group sensory therapy for older
  adults with dementia. In: 2013 IEEE 13th International Conference on
  Rehabilitation Robotics (ICORR). pp.~1--6. IEEE (2013)

\bibitem{santini2020social}
Santini, Z.I., Jose, P.E., Cornwell, E.Y., Koyanagi, A., Nielsen, L.,
  Hinrichsen, C., Meilstrup, C., Madsen, K.R., Koushede, V.: Social
  disconnectedness, perceived isolation, and symptoms of depression and anxiety
  among older americans (nshap): a longitudinal mediation analysis. The Lancet
  Public Health  \textbf{5}(1),  e62--e70 (2020)

\bibitem{Scassellatieabc9014}
Scassellati, B., V{\'a}zquez, M.: The potential of socially assistive robots
  during infectious disease outbreaks. Science Robotics  \textbf{5}(44) (2020).
  \doi{10.1126/scirobotics.abc9014},
  \url{https://robotics.sciencemag.org/content/5/44/eabc9014}

\bibitem{share2018preparing}
Share, P., Pender, J.: Preparing for a robot future? social professions, social
  robotics and the challenges ahead. Irish Journal of Applied Social Studies
  \textbf{18}(1), ~4 (2018)

\bibitem{shibata2012therapeutic}
Shibata, T.: Therapeutic seal robot as biofeedback medical device: Qualitative
  and quantitative evaluations of robot therapy in dementia care. Proceedings
  of the IEEE  \textbf{100}(8),  2527--2538 (2012)

\bibitem{smith2002psychosocial}
Smith, T.W., Ruiz, J.M.: Psychosocial influences on the development and course
  of coronary heart disease: current status and implications for research and
  practice. Journal of consulting and clinical psychology  \textbf{70}(3), ~548
  (2002)

\bibitem{usher2020family}
Usher, K., Bhullar, N., Durkin, J., Gyamfi, N., Jackson, D.: Family violence
  and covid-19: Increased vulnerability and reduced options for support.
  International journal of mental health nursing  (2020)

\bibitem{usher2020life}
Usher, K., Bhullar, N., Jackson, D.: Life in the pandemic: Social isolation and
  mental health. Journal of Clinical Nursing  (2020)

\bibitem{van2020using}
Van~Bavel, J.J., Baicker, K., Boggio, P.S., Capraro, V., Cichocka, A., Cikara,
  M., Crockett, M.J., Crum, A.J., Douglas, K.M., Druckman, J.N., et~al.: Using
  social and behavioural science to support covid-19 pandemic response. Nature
  Human Behaviour pp. 1--12 (2020)

\bibitem{wada2005psychological}
Wada, K., Shibata, T., Saito, T., Sakamoto, K., Tanie, K.: Psychological and
  social effects of one year robot assisted activity on elderly people at a
  health service facility for the aged. In: Proceedings of the 2005 IEEE
  international conference on robotics and automation. pp. 2785--2790. IEEE
  (2005)

\bibitem{wada2002robot}
Wada, K., Shibata, T., Saito, T., Tanie, K.: Robot assisted activity for
  elderly people and nurses at a day service center. In: Proceedings 2002 IEEE
  International Conference on Robotics and Automation (Cat. No. 02CH37292).
  vol.~2, pp. 1416--1421. IEEE (2002)

\bibitem{wang2017robots}
Wang, R.H., Sudhama, A., Begum, M., Huq, R., Mihailidis, A.: Robots to assist
  daily activities: views of older adults with alzheimer's disease and their
  caregivers. International psychogeriatrics  \textbf{29}(1),  67--79 (2017)

\bibitem{yang2020combating}
Yang, G.Z., Nelson, B.J., Murphy, R.R., Choset, H., Christensen, H., Collins,
  S.H., Dario, P., Goldberg, K., Ikuta, K., Jacobstein, N., et~al.: Combating
  covid-19?the role of robotics in managing public health and infectious
  diseases (2020)

\bibitem{zeng2020high}
Zeng, Z., Chen, P.J., Lew, A.A.: From high-touch to high-tech: Covid-19 drives
  robotics adoption. Tourism Geographies pp. 1--11 (2020)

\bibitem{Paro}
{Šabanović}, S., {Bennett}, C.C., {Chang}, W., {Huber}, L.: Paro robot
  affects diverse interaction modalities in group sensory therapy for older
  adults with dementia. In: 2013 IEEE 13th International Conference on
  Rehabilitation Robotics (ICORR). pp.~1--6 (2013)

\end{thebibliography}

\end{document}